\documentclass[preprint,12pt]{elsarticle}
\usepackage{amsmath}
\usepackage{amssymb}
\usepackage[pdftex,colorlinks]{hyperref}
\usepackage{multirow}

\biboptions{compress}

\begin{document}

\begin{frontmatter}

\title{Dissipation attack on Bennett-Brassard 1984 protocol in practical quantum key distribution system}

\author[a]{Li Yang\corref{1}}\ead{yangli@iie.ac.cn}
\author[b]{Bing Zhu}
\cortext[1]{Corresponding author.}
\address[a]{State Key Laboratory of Information Security, Institute of Information Engineering, Chinese Academy of Sciences, Beijing 100093, China}
\address[b]{Department of Electronic Engineering and Information Science, University of Science and Technology of China, Hefei, Anhui 230027, China}

\begin{abstract}
We propose a new kind of individual attack, based on randomly selected dissipation, on Bennett-Brassard 1984 protocol of practical quantum key distribution (QKD) system with lossy and noisy quantum channel.
Since an adversary with super quantum channel can disguise loss and errors induced by his attack as that of the system, he can obtain innegligible amount of information for a practical QKD system, without being detected by legal participants.

\end{abstract}


\end{frontmatter}

% \linenumbers

The dissipation of quantum channel has been employed in attacking Bennett 1992 protocol in a practical quantum key distribution system\cite{Yan03}. Here we propose a new kind of individual attack named dissipation attack to get information of the row key generated by Bennett-Brassard 1984 protocol in a practical QKD system with lossy and noisy quantum channel.

\noindent{\emph{\textbf{1.} Analysis of dissipation along the direction of a linear polarized photon state:}} as shown in $Fig. 1$, $a, b$ are two orthogonal directions,
any optical signal will be decomposed into two components along $a, b$ direction by polarization beam splitter. Suppose that light intensity of the signal is $I$, the intensity of components
 in directions $a, b$ are $I_a, I_b$ and the amplitudes are $A_0, A'_0$, respectively. Assume that the direction of dissipation is $a$,
 the ratio of intensity dissipation is $\alpha$, the amplitude in direction $a$  after dissipation is $A$, then we have %so $I=I_a+I_b$£¬$I_a\propto A_0^2$,

\begin{eqnarray}
  \frac{A}{A_0}=\sqrt{\alpha},
\end{eqnarray}
thus
\begin{eqnarray}
 \frac{tg\theta}{tg\theta_0}=\sqrt{\alpha}.
\end{eqnarray}

Since the polarization direction of transmitted photon has been changed from $\theta_0$ to $\theta$, it is clear that the attack dissipation will lead to errors.

\begin{figure}[htbp]
 \centering
\includegraphics[scale=0.4,bb= 0 0 440 233]{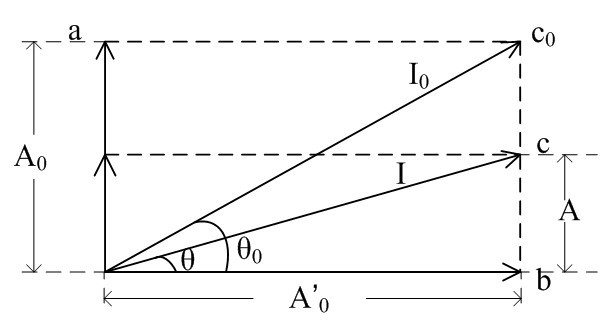}\\
  \caption{Dissipation along the polarization direction of a polarized qubit}\label{pic1}
\end{figure}

\noindent{ \emph{\textbf{2.} Randomly selected dissipation attack}:} Consider the quantum key distribution system using Bennett-Brassard 1984 protocol, which transmitting photon in states $|H\rangle,|\frac{\pi}{4}\rangle$ for 0, and $|V\rangle,|\frac{3\pi}{4}\rangle$ for 1.
\begin{figure}[htb]
 \centering
\includegraphics[scale=0.3,bb= 0 0 376 237]{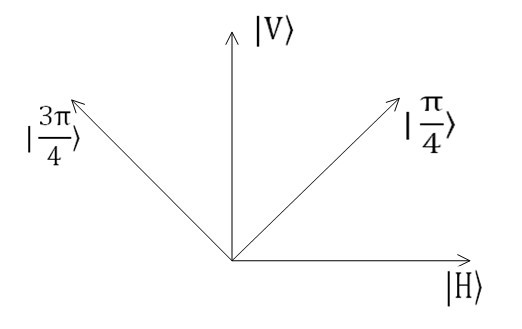}\\
  \caption{Conjugate coding of photon state }\label{pic0}
\end{figure}

Our attack scheme is as follows. The adversary randomly selects each of the four polarization directions as the dissipation direction to keep the balance between 0 and 1 of the bit string transmitted in the quantum channel, so that the legal users will not find the existence of the adversary. It is necessary to assume that the adversary has a super quantum channel with much less loss and error.

We shall also investigate the attack scheme with dissipation randomly along one of the two directions of Breidbart bases\cite{Ben82}.

{ (1) \emph{Attack with dissipation along the polarization direction of state $|V\rangle$}.}
If we chose the dissipation direction along the polarized direction of one of the four states of BB84 protocol, the attack effect is the same. So, without loss generality, we choose the direction of $|V\rangle$ as dissipation direction to analysis the attack effect. For qubits in states $|H\rangle$, $|V\rangle$, intensity of the qubits will be changed, though such change does not cause measurement errors. However, for qubits in states $|\frac{\pi}{4}\rangle$, $|\frac{3\pi}{4}\rangle$, errors will occur after attack, only the components in the direction $|V\rangle$ will dissipate.

\begin{figure}[htbp]
 \centering
\includegraphics[scale=0.4,bb= 0 0 612 265]{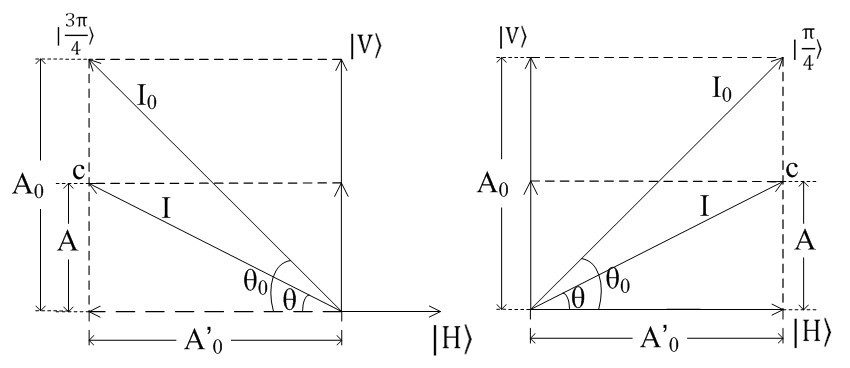}\\
  \caption{Dissipation scheme 1}\label{pic2}
\end{figure}

We define the bit enduring large dissipation the strong dissipation bit $b$. In contrast, weak dissipation bit is denoted as $\bar{b}$. We can see that after dissipation with ratio $\alpha$, the intensity of a strong dissipation bit reduces to

\begin{eqnarray}
  I_b^{(1)}=\frac{1}{2}(I_{|V\rangle}+I_{|\frac{3\pi}{4}\rangle})=\frac{1}{2}(\alpha I_0+\frac{1}{2}\alpha I_0+\frac{1}{2}I_0)=\frac{1}{4}(1+3\alpha)I_0,
\end{eqnarray}
and the intensity of a weak dissipation bit reduces to
\begin{eqnarray}
  I_{\bar{b}}^{(1)}=\frac{1}{2}(I_{|H\rangle}+I_{|\frac{\pi}{4}\rangle})=\frac{1}{2}(I_0+\frac{1}{2}\alpha I_0+\frac{1}{2}I_0)=\frac{1}{4}(3+\alpha)I_0,
\end{eqnarray}
then the average intensity of each bit decreases to£º
\begin{eqnarray}\label{(2)}
  I^{(1)}=\frac{1}{2}(I_b^{(1)}+I_{\bar{b}}^{(1)})=\frac{1}{2}(1+\alpha)I_0.
\end{eqnarray}
The probability of a legal user receiving the strong dissipation bit $b$ is:

\begin{eqnarray}\label{2a}
  p_b^{(1)}=\frac{I_b^{(1)}}{I_b^{(1)}+I_{\bar{b}}^{(1)}}=\frac{1}{2}-\frac{1}{4}\times\frac{1-\alpha}{1+\alpha},
\end{eqnarray}
the probability of receiving the weak dissipation bit $\bar{b}$ is£º
\begin{eqnarray}\label{2b}
  p_{\bar{b}}^{(1)}=\frac{I_{\bar{b}}^{(1)}}{I_b^{(1)}+I_{\bar{b}}^{(1)}}=\frac{1}{2}+\frac{1}{4}\times\frac{1-\alpha}{1+\alpha},
\end{eqnarray}
where $p_b^{(1)}+p_{\bar{b}}^{(1)}=1$, $\frac{1}{4}\leq p_b^{(1)}\leq\frac{1}{2}\leq p_{\bar{b}}^{(1)}\leq\frac{3}{4}$.

Since the polarization of states $|\frac{\pi}{4}\rangle$,$|\frac{3\pi}{4}\rangle$ has been changed during the dissipation, they will lead to errors. For $\theta_0=\frac{\pi}{4}$, we obtains error rate as below:

\begin{eqnarray}
  \eta^{(1)}=\frac{1}{2}\sin^2(\theta_0-\theta)=\frac{1}{4}\times\frac{1-\alpha}{1+\alpha}\times\frac{1-\sqrt{\alpha}}{1+\sqrt{\alpha}}.
\end{eqnarray}\\

{(2) \emph{Dissipation attack along the direction of Breidbart bases.}}
\begin{figure}[htbp]
 \centering
\includegraphics[scale=0.4,bb= 0 0 384 256]{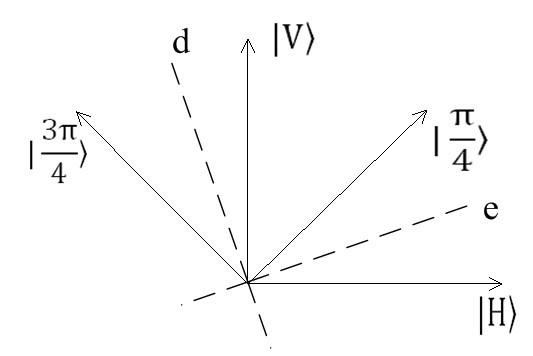}\\
  \caption{Breidbart bases}\label{pic3}
\end{figure}

If dissipation direction is one of $d$ and $e$ which are used in  Breidbart attack on the BB84 protocol, the effect of dissipation is the same, so we select $d$ as dissipation direction to analysis the effect. Intensity of the sending states will all be effected. In this situation, intensity of the bit $b=1$ will be effected most by dissipation, so we call $b=1$ as strong dissipation bit and $\bar{b}=0$ as weak dissipation bit.

Suppose the dissipation has been executed on every bit. After dissipation with ratio $\alpha$, the intensity change of strong dissipation bit is shown on the right in $Fig. 5$, and that of weak dissipation bit is on the left. We now take qubit in states $|V\rangle, |\frac{\pi}{4}\rangle$ as example to show the effect of dissipation. It can be seen that the dissipation effect of qubit in states $|\frac{3\pi}{4}\rangle,|H\rangle$ is the same as that in $|V\rangle,|\frac{\pi}{4}\rangle$.

\begin{figure}[htbp]
 \centering
\includegraphics[scale=0.4,bb= 0 0 718 334]{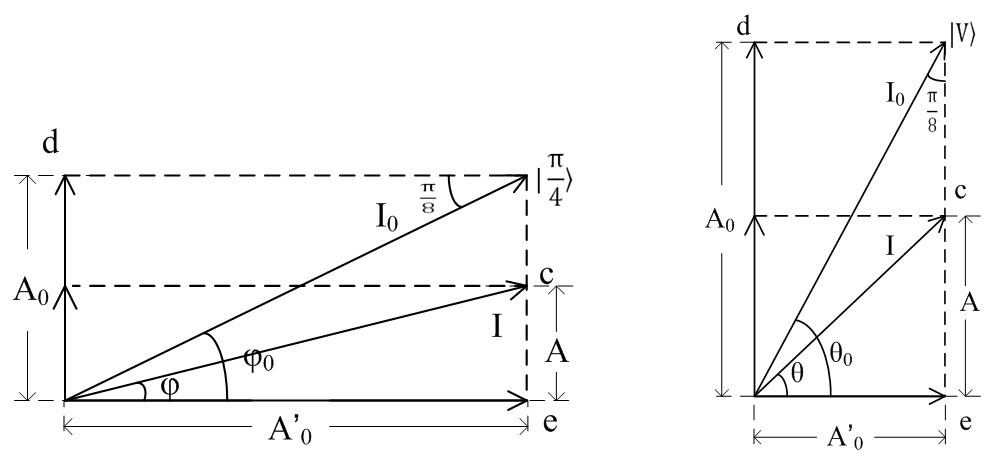}\\
  \caption{Dissipation scheme 2}\label{pic4}
\end{figure}

By dissipation in the direction $d$, the intensity of the strong dissipation bit corresponding to qubit in state $|V\rangle$ or state $|\frac{3\pi}{4}\rangle$ reduces to

\begin{eqnarray}
  I_b^{(2)}=\frac{1}{2}(I_{|V\rangle}+I_{|\frac{3\pi}{4}\rangle})=(\alpha \cos^2(\frac{\pi}{8})+\sin^2(\frac{\pi}{8}))I_0=\frac{\alpha+(\sqrt{2}-1)^2}{1+(\sqrt{2}-1)^2}I_0,
\end{eqnarray}
and the intensity of the  weak dissipation bit reduces to
\begin{eqnarray}
  I_{\bar{b}}^{(2)}=\frac{1}{2}(I_{|H\rangle}+I_{|\frac{\pi}{4}\rangle})=(\alpha \sin^2(\frac{\pi}{8})+\cos^2(\frac{\pi}{8}))I_0=\frac{1+\alpha(\sqrt{2}-1)^2}{1+(\sqrt{2}-1)^2}I_0,
\end{eqnarray}
then the average intensity becomes
\begin{eqnarray}
  I^{(2)}=\frac{1}{2}(I_b^{(1)}+I_{\bar{b}}^{(1)})=\frac{1}{2}(1+\alpha)I_0.
\end{eqnarray}
The probability of a legal user receiving a strong dissipation bit $b$ is

\begin{eqnarray}\label{3a}
  p_b^{(2)}=\frac{I_b^{(1)}}{I_b^{(1)}+I_{\bar{b}}^{(1)}}=\frac{1}{2}-\frac{\sqrt{2}}{4}\times\frac{1-\alpha}{1+\alpha},
\end{eqnarray}
and the probability of a legal user receiving a weak dissipation bit $\bar{b}$ is

\begin{eqnarray}\label{3b}
  p_{\bar{b}}^{(2)}=\frac{I_{\bar{b}}^{(1)}}{I_b^{(1)}+I_{\bar{b}}^{(1)}}=\frac{1}{2}+\frac{\sqrt{2}}{4}\times\frac{1-\alpha}{1+\alpha}.
\end{eqnarray}
Where $p_b^{(2)}+p_{\bar{b}}^{(2)}=1$, $\frac{2-\sqrt{2}}{4}\leq p_b^{(2)}\leq\frac{1}{2}\leq p_{\bar{b}}^{(2)}\leq\frac{2+\sqrt{2}}{4}$.

In this kind of attack, error occurs in the measurement of all the four states. For the strong dissipation bit, $\theta_0=\frac{3\pi}{8}, tg\theta=(\sqrt{2}+1)\sqrt{\alpha}$, error rate is

\begin{eqnarray}
  \eta_b^{(2)}=\sin^2(\theta_0-\theta)=\frac{(1-\sqrt{\alpha})^2}{(\sqrt{2}-1+(\sqrt{2}+1)\sqrt{\alpha})^2+(1-\sqrt{\alpha})^2},
\end{eqnarray}
For the weak dissipation bit, $\varphi_0=\frac{\pi}{8}, tg\varphi=(\sqrt{2}-1)\sqrt{\alpha}$, error rate is
\begin{eqnarray}
  \eta_{\bar{b}}^{(2)}=\sin^2(\varphi_0-\varphi)=\frac{(1-\sqrt{\alpha})^2}{(\sqrt{2}+1+(\sqrt{2}-1)\sqrt{\alpha})^2+(1-\sqrt{\alpha})^2},
\end{eqnarray}
So in this attack scheme, average error rate is

\begin{eqnarray}
  \eta^{(2)}=\frac{1}{2}(\eta_b^{(2)}+\eta_{\bar{b}}^{(2)})=\frac{1}{4}\times\frac{(1-\sqrt{\alpha})^2}{1+\alpha}.
  \end{eqnarray}
It can be seen that
\begin{eqnarray}
  \eta^{(2)}=\eta^{(1)},
\end{eqnarray}
so the error rate induced by these two dissipation attack schemes can be denoted by one parameter $\eta$, and 

\begin{eqnarray}\label{(4)}
  \eta=\frac{1}{4}\times\frac{1-\alpha}{1+\alpha}\times\frac{1-\sqrt{\alpha}}{1+\sqrt{\alpha}}.
\end{eqnarray}\\

\noindent{  \emph{\textbf{3.} Relation between error rate and the information adversary obtained.}} For the first attacking scheme, we have

\begin{eqnarray}
  \alpha=\frac{4p_b^{(1)}-1}{3-4p_b^{(1)}},
\end{eqnarray}
then,

\begin{eqnarray}\label{(5)}
  \eta=\frac{1}{2}\left(\sqrt{\frac{3}{4}-p_b^{(1)}}-\sqrt{p_b^{(1)}-\frac{1}{4}}\right)^2.
\end{eqnarray}

For the second attacking scheme, we have

\begin{eqnarray}
  \alpha=\frac{1-\sqrt{2}(1-2p_b^{(2)})}{1+\sqrt{2}(1-2p_b^{(2)})},
\end{eqnarray}
then,
\begin{eqnarray}\label{(6)}
  \eta=\frac{\sqrt{2}}{4}\left(\sqrt{\frac{2+\sqrt{2}}{4}-p_b^{(2)}}-\sqrt{p_b^{(2)}-\frac{2-\sqrt{2}}{4}}\right)^2.
\end{eqnarray}
\begin{figure}[htb]
 \centering
\includegraphics[scale=0.4,bb= 200 0 695 260]{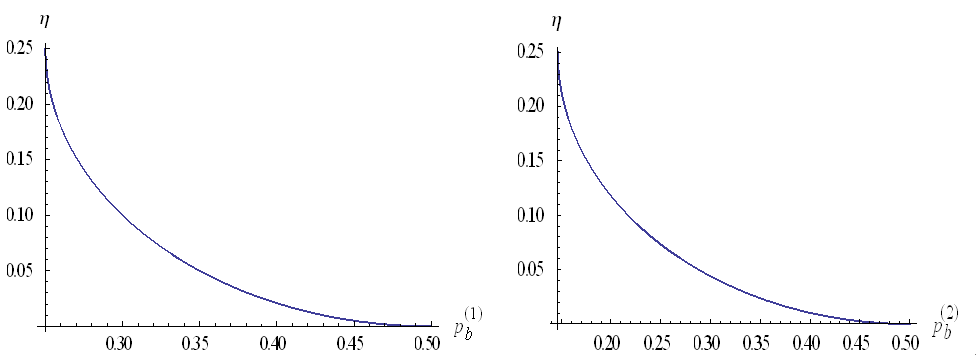}\\
  \caption{Relations between probabilities the adversary get and the error rate caused by the attacks}\label{pic5}
\end{figure}
The relations between $\eta$ and $p^{(i)}_b$ $(i=1,2)$ are shown in $Fig. 6$.

Let the adversary's information of $i_{th}$ attack scheme is $H_i=1-H(p_b^{(i)})$. The relations between error rate $\eta$ and $H_{i}, i=1,2$ shown in $Fig. 7$.
\begin{figure}[htb]
 \centering
\includegraphics[scale=0.4,bb= 0 0 474 313]{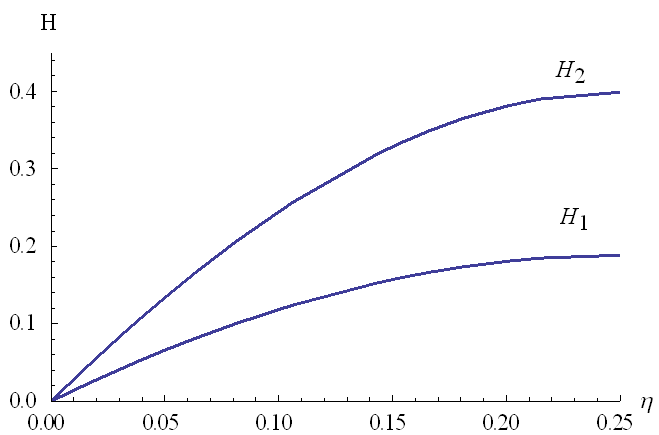}\\
  \caption{Relations between error rate caused by attacks and the information obtained by adversary}\label{pic6}
\end{figure}
Though our schemes are not belong to general individual attack based on probe qubits and unitary transformation, they still satisfy an inequality presented in \cite{Gis02}: $H/\eta < 2.9$. Why this relation still holds for the dissipation attack is worth to be investigated.

\noindent{\emph{\textbf{4.} Analysis of feasibility}:} assume the adversary own a supper quantum channel without loss and errors. For simplicity, assume the legal user have an ideal single photon detector. In the first attack, the adversary randomly chooses four directions to dissipate, so the user will generate a uniform distribution ``0,1" bit string, meanwhile, according to the adversary, the bit string dissipated with $|H\rangle,|\frac{\pi}{4}\rangle$'s direction will generate 1 with greater probability, as who dissipated with $|V\rangle,|\frac{3\pi}{4}\rangle$'s direction will generate 0 with greater probability.

In order to avoid being caught, it is required that not only the ``0, 1" distribution of the bit string tempered is uniform, but also the loss and error rate induced should equal to the system's intrinsic loss and error rate. As long as the system's error rate reach $5\%$, the dissipation rate $\alpha$ can reach $0.251$, it corresponds to the loss of 10.2 km standard single-mode fiber. Meanwhile, the effect of attack is $p_b^{(1)}=0.350,p_b^{(2)}=0.288$, and the adversary obtains $H_1=0.066, H_2=0.133$ bit information each bit of the key, respectively.

The attacks described above does not require the adversary equipped with single-photon source, single-photon detector, probe qubits and related unitary evolution. It will be accomplished if the adversary has super quantum channel and can dissipate each qubit in a randomly chosen direction. By comparison, photon-number-splitting attack and intercept-resend attack require more abilities for the adversary.

\section*{Acknowledgment}
This work was supported by National Natural Science Foundation of China under Grant No. 61173157.

\bibliographystyle{model1a-num-names}
\bibliography{<your-bib-database>}

\begin{thebibliography}{00}
\bibitem{Yan03}L. Yang, L. -A. Wu, e-print arXiv: quant-ph/0310080, 2003.
\bibitem{Ben92}C. H. Bennett, Phys. Rev. Lett., {\bf68}, 3121(1992).
\bibitem{Ben84}C. H. Bennett and G. Brassard, in {\sl Proceedings of the IEEE International Conference on Computers, Systems and signal Processing} (IEEE, New York), p. 175 (1984).
\bibitem{Eke94}A. K. Ekert, B. Huttner, et al.  Phys. Rev. {\bf A 50}(2), pp.1047-1056 (1994).
\bibitem{Gis02}N. Gisin, G. Ribordy, W. Tittel, and H. Zbinden, Rev. Mod. Phys., {\bf74}(1), pp.145-195(2002).
\bibitem{Ben82}C. H. Bennett, G. Brassard, S. Breidbart, and S. Wiesner, In Advances in Cryptology¨CProceedings of Crypto'82, pp.267-275 (1983).
\end{thebibliography}



%% References

%%\bibliographystyle{model1a-num-names}
%%\bibliography{<your-bib-database>}





% \end{thebibliography}
%\end{CJK*}
\end{document}